\documentclass[]{raa}            
\usepackage{graphicx,times}
\usepackage{natbib}

\providecommand{\bm}[1]{\mbox{\boldmath$#1$\unboldmath}}

\def\micron{\hbox{$\mu$m}}

\begin{document}

   \title{Near-Infrared Imaging Survey of Faint Companions around 
   Young Dwarfs in the Pleiades Cluster
 $^*$
\footnotetext{\small $*$ Based on data collected at the Subaru Telescope,
which is operated by the National Astronomical Observatory of Japan.}
}

 \volnopage{ {\bf 2009} Vol.\ {\bf 9} No. {\bf XX}, 000--000}
   \setcounter{page}{1}

   \author{Yoichi Itoh
      \inst{1}
   \and Yumiko Oasa
      \inst{2}
   \and Hitoshi Funayama
      \inst{1}
   \and Masahiko Hayashi
      \inst{3,4}
   \and Misato Fukagawa
      \inst{5}
   \and Toshio Hashiguchi
      \inst{1}
   }

   \institute{Graduate School of Science, Kobe University, 
             1-1 Rokkodai, Nada-ku, Kobe 657-8501, Japan;
             {\it yitoh@kobe-u.ac.jp}\\
        \and
             Faculty of Education, Saitama University,
             255 Shimo-Okubo, Sakura-ku, Saitama, Saitama 338-8570, Japan
        \and
             Subaru Telescope,
             650 North A`oh\=ok\=u Pl., Hilo, Hawaii 96720, USA
        \and
             School of Mathematical and Physical Science, Graduate University
             for Advanced Studies (Sokendai),
             2-21-1 Osawa, Mitaka, Tokyo 181-8588, Japan
        \and
             Graduate School of Science, Osaka University,
             Machikaneyama, Toyonaka, Osaka 560-0043, Japan
\vs \no
   {\small Received [year] [month] [day]; accepted [year] [month] [day] }
}

\abstract{
We conducted a near-infrared imaging survey of 11 young dwarfs 
in the Pleiades cluster 
using the Subaru Telescope and the near-infrared coronagraph imager.
We found 10 faint point sources, with magnitudes as
faint as 20 mag in the $K$-band, around 7 dwarfs.
Comparison with Spitzer archive images revealed that a pair
of the faint sources around V 1171 Tau are very red in the infrared
wavelengths, indicative of very low-mass young stellar objects.
However, the results of our follow-up proper motion measurements implied that
the central star and the faint sources do not share common proper motions,
suggesting that they are not physically associated.
}

   \authorrunning{Y. Itoh et al.}            
   \titlerunning{Near-Infrared Survey of Companions in Pleiades}  
   \maketitle

%
%
\section{Introduction}
\label{sect:intro}

Extrasolar planet searches have located about 300 planet candidates.
Their physical characteristics differ significantly from those in our Solar system.
Many systems have gas giants orbiting close to their central stars;
in other cases, planetary orbits are highly eccentric.
The current sample of extrasolar planets, however, is biased toward planets
with small orbital radii:
more than half have semimajor axes smaller than 1 AU.
This stems from using detection methods 
that possess increased sensitivity to objects orbiting closer to the central stars.
A comprehensive view of extrasolar planetary systems, including bodies with large orbital radii,
has yet to be completed.

Direct imaging of extrasolar planets is more sensitive to planets with 
larger orbital radii.
The dynamic range of imaging required to detect a gas giant planet is extremely high:
10$^{9}$ for Jupiter against the Sun in the optical and near-infrared wavelengths.
This range, however, is significantly reduced if a planetary system is young.
Such young systems are located relatively far away ($\sim$100~pc) compared 
to the more evolved stars in the solar neighborhood ($\sim$10~pc).
These young systems are still detectable with an 8--10~m class telescope 
combined with an adaptive optics (AO) system, provided the orbital radii are larger than several tens of AU.
In fact, detections of marginal planet candidates
or planet-brown dwarf boundary-mass objects were reported 
in some favorable cases (e.g., DH Tau, Itoh et al. 2005; 2M1207, Chauvin et al. 2005a; GQ Lup, Neuh\"auser et al. 2005;
AB Pic, Chauvin et al. 2005b; CHXR 73, Luhman et al. 2006; RXS J160929.1-210524, Lafreniere et al. 2008).
In addition, very recently
Marois et al. (2008) and Kalas et al. (2008) separately obtained direct
images of extrasolar planets around young A-type stars, HR 8799 and Fomalhaut.

In this paper, we present the results of a near-infrared imaging survey of 
faint companions around 11 young dwarfs in the Pleiades cluster.
The age of Pleiades is 115 Myr (Basri et al. 1996); thus
planets, as well as stars, are expected to be in the bright contraction phase.

\section{Observations and Data Reduction}
\label{sect:obs}

We selected single stars as targets, as
the region in which a planet survives is restricted in a multiple-star
system (Holman \& Wiegert 1999).
We chose objects that were
confirmed as single stars by at least two of the following methods:
1) adaptive optics observations (Bouvier et al. 1997), 
2) Doppler shift measurements (Mermilliod et al. 1992), and
3) photometric measurements (Kaehler 1999).
We then checked membership
probability and used only candidates whose probability was 
greater than 70\% based on proper-motion
measurements (Schilbach et al. 1995; Deacon \& Hambl 2004).

The $H$-band observations were conducted on 2004 Dec 28 and the $K$-band imaging
on 2005 Nov 17 (table \ref{log}).
The near-infrared coronagraph camera CIAO (Tamura et al. 2000; Murakawa et al. 2004) mounted 
on the Subaru Telescope was combined with the 36-actuator AO system (Takami et al. 2004).
CIAO is equipped with a 1024$\times$1024 InSb Alladin~II detector with a spatial scale 
of 0\farcs0213 pixel$^{-1}$.
The spatial resolution provided by the AO system was 0\farcs1--0\farcs2 (FWHM) 
depending on the natural seeing of 0\farcs5 to 1\arcsec at the time of observations.
The single stars themselves were used as the AO reference stars.
Most single stars were occulted by a 0\farcs6-diameter
mask constructed of a sapphire substrate coated with chromium, which provided
a transmittance of a few tenths of a percent.
This allowed us to measure accurate positions for the occulted stars.
We used a traditional circular Lyot stop with an 80~\% diameter of the pupil.
Due to instrument problems, background noise was extremely large during
the $H$-band observations.
During the $K$-band observations, occasional clouds appeared and the seeing was poor.
For each object, the telescope pointing was finely adjusted so that 
the target star was placed at the center of the occulting mask. 
Each exposure time was 10 s,
and the total integration time varied from 40 s to 720 s,
depending on the brightness of the central star and the weather conditions.
After several frames were taken, 
both the telescope and the occulting mask were dithered by $\sim1\arcsec$.
Then, the same target was again placed at the center of the occulting mask, 
and additional frames were acquired.
Faint standard stars (FSs; Hawarden et al. 2001) were observed for photometric calibration.
Dark frames and dome flats with incandescent lamps were taken at the end of each night.

The follow-up imaging observation of V 1171 Tau was conducted on 2009 Aug 15.
We used the IRCS instrument combined with the new AO 188 instrument.
We did not use any occulting mask.
The observations were taken in the $K$-band.
Fifteen frames were obtained with 0.5 s integration time and 25 frames 
with  10 s integration time.
The FWHM of the PSFs was 0\farcs08.

The Image Reduction and Analysis Facility (IRAF) was used for data reduction.
After a dark frame was subtracted from each object frame, each
object frame was divided 
by the dome flat, and hot and bad pixels were removed.  
In addition to the standard data reduction process described above, 
we applied the following method to detect faint companions around the central bright stars:
we removed the halo of the central star in each frame by subtracting 
the median-filtered image of the object itself, 
rather than subtracting separate images
of a point spread function (PSF) reference star, 
as we were interested in ``point-like sources.''
We created the median-filtered image using the FRMEDIAN task,
which produces a circular median-filtered image.
A large-radius median filter conserves the flux of a faint point
source, but weakly suppresses the halo of the bright central star.
We set the radius of the median window to twice the FWHM of the
point source in the image.
The peak position of the central star moved slightly on the detector during the observations.
This was due to the difference in atmospheric dispersion between the infrared wavelength, 
at which the images were taken, and the optical wavelength, in which the wavefront sensing was applied.
We compensated for this effect by shifting the image frame so that the peak position 
of the central star coincided with the frame center, after the peak position was measured 
with the RADPROFILE task in the IRAF.
The halo of the star was suppressed in each frame 
after the median-filtered image was subtracted
and all of the frames for each target were combined into the final image.

To detect faint sources, we used the median filtered images.
We used the SExtractor program with a 3$\sigma$-detection threshold above the background  
to detect companion candidates.
We excluded objects with PSF ellipticities larger 
than 0.25 or semiminor axes smaller than 1.3 pixels.
As described below, all detected faint sources were located far from the bright 
star.
To conserve the flux of the faint source, we measured the flux
in the image without first subtracting the PSF of the central star.
We measured the object's brightness by aperture photometry using the APPHOT
task.
Since the seeing was poor and some objects seemed to be extended,
we used several aperture radii (table 2).
For V1171 Tau, due to the small separation between cc1 and cc2,
we could not use an appropriate radius aperture.
Instead, we made our photometric measurement 
using a small aperture (8 pixels) and then
applied the aperture correction.
We determined the amount of the aperture correction 
by comparing a measurement of 
V1171 Tau/cc4 using the small aperture to a measurement made with an appropriately wide aperture (15 pixels).
For HD 23269, we measured the photometry of the companion candidate relative
to the central star.
In general, PSFs corrected by the AO system are often degraded, in particular
by anisoplanatic elongation.
This may cause an additional systematic uncertainty in the photometric results.
We confirmed that the ellipticities of the PSFs
were less than 0.15 for all candidates but extended objects.
By measuring the objects with different aperture radii, we estimated that the
systematic uncertainties in the photometry were about 0.2 mag.

\section{Results and Discussion}
\label{sect:res}

\begin{figure}[h!!!]
\centering
\includegraphics[width=14.0cm, angle=0]{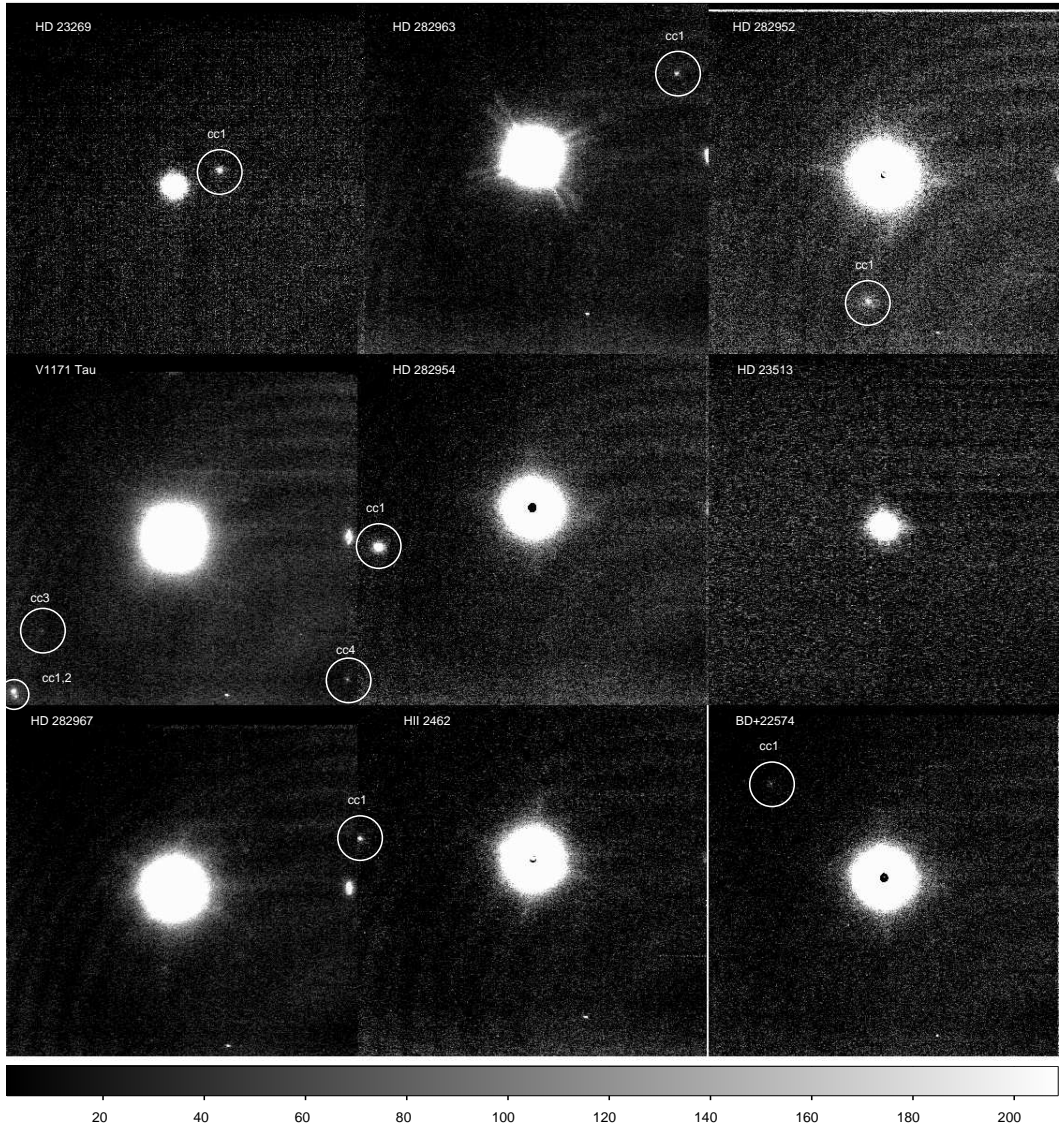}
\begin{minipage}[]{140mm}
\caption{ 
$K$-band images of the observed objects.
The central star PSFs are not suppressed in these images.
The objects marked with ``cc'' are companion candidates.
The other structures, aside from the central stars, are ghosts and hot pixels.
The structures seen in all images with the same separations and position angles
relative to the central star are identified as ghosts.
The structures seen at the same pixels in all images are identified
as hot pixels.
The field of view for each is 20\arcsec$\times$20\arcsec.
\label{allfig}
}
\end{minipage}
\end{figure}

Ten faint sources were detected around 7 dwarfs
of the 11 dwarfs surveyed (figure \ref{allfig} and table 2).
We assumed a completeness limit around 20 mag at the $K$-band,
although this is not uniform through the survey. 
The number of background stars not associated with the Pleiades cluster
was then estimated from the star-count 
model of the Galaxy (Jones et al. 1981),
giving an average of 0.62 background stars within the limiting magnitude 
in the CIAO field of view ($22\arcsec \times 22\arcsec$),
or 7 for the entire survey.
We assert that the detected number of the faint sources is larger than 
expected from a random distribution, although the significance is at most at
the 1$\sigma$-level.
This prompted us to infer that some of the detected sources in our survey were
not merely background stars, 
but actual faint objects located in the Pleiades cluster.

\begin{figure}[h!!!]
\centering
\includegraphics[width=14.0cm, angle=0]{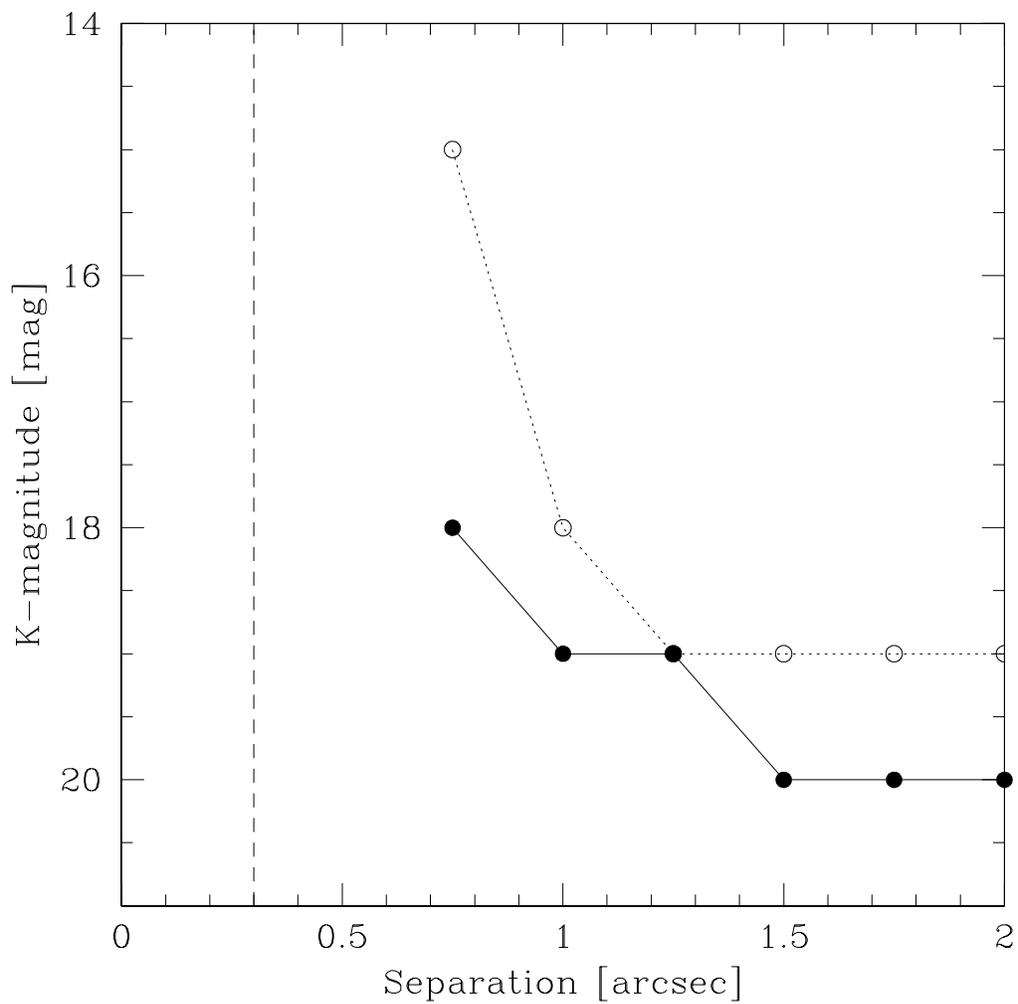}
\begin{minipage}[]{140mm}
\caption{ 
Limiting magnitudes of the observation (filled circles: V1171 Tau; 
open circles: HD282954). 
The limiting magnitudes are estimated by adding pseudostars. 
This observation is, to date, the deepest search for extrasolar 
planets and brown dwarfs around young dwarfs in the Pleiades cluster.
\label{addstar}
}
\end{minipage}
\end{figure}

We estimated the detection limit for a point source by adding pseudo-PSFs 
to the raw data. 
We defined Gaussian PSFs with 12 to 22 mag at 1 mag intervals. 
Their FWHMs are the same as that of the faint sources in the image. 
We placed them between 0\farcs25 and 1\farcs5 with a 0\farcs25 interval, 
as well as between 2\farcs0 and 10\farcs0 with a 1\farcs0 
interval from the central star. 
At each separation, the pseudo-PSFs are located 
at P.A. = --90\fdg0, --45\fdg0, 0\fdg0, and +45\fdg0. 
When three or four PSFs at the same separation are identified by 
the SExtractor program, the object is classified as detected. 
The limiting magnitude is shown in figure \ref{addstar}
as a function of separation from the central star. 
Beyond 1\farcs25 from the central star, 
the limiting magnitude is as deep as 20 mag at the $K$-band, 
corresponding to a 10 $M_{\rm J}$ planet at the same age of the Pleiades member.
In the region within 1\arcsec~ from the central star, 
detection sensitivity is severely restricted by the residual halo 
of the central star.

\subsection{V1171 Tau}

A $K$-band image of V1171 Tau is presented in the middle-left panel of
figure \ref{allfig}.
Four point sources, whose
$K$-band magnitudes are as faint as 20 mag, appear around the central star.
If associated with the central star, these
are very low-mass, or planetary-mass, companions.

\begin{figure}[h!!!]
\centering
\includegraphics[width=14.0cm, angle=0]{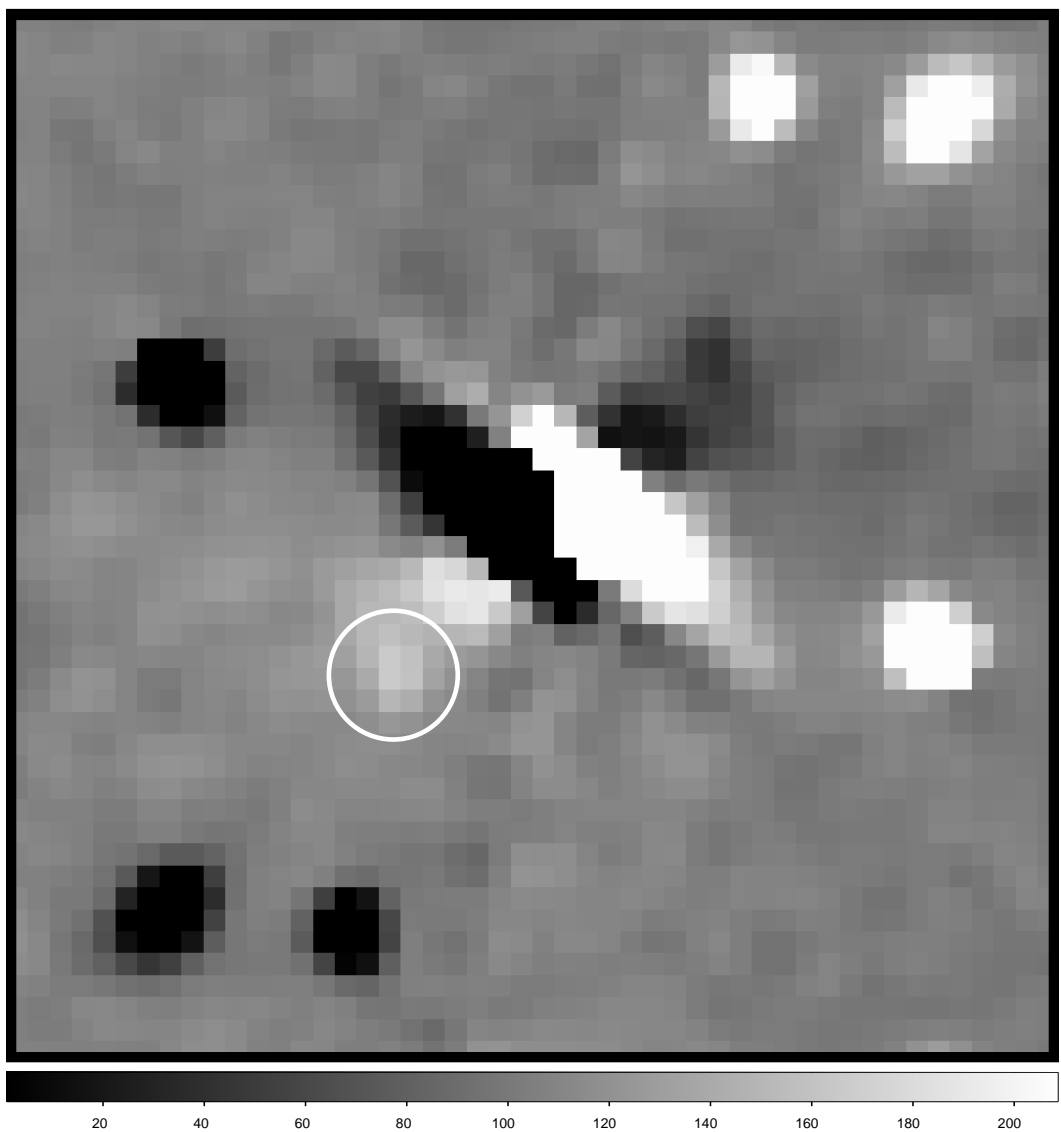}
\begin{minipage}[]{140mm}
\caption{ 
3.6\micron~ Spitzer image of V1171 Tau.
By subtracting a 180 \degr~ rotated image,
the flux of the bright central star is deeply suppressed.
The companion candidates cc1 and cc2 are marked by a circle,
but not spatially resolved.
The field of view is 60\arcsec$\times$60\arcsec.
\label{Spitzer}
}
\end{minipage}
\end{figure}

V 1171 Tau/cc1 and cc2 seem to form a binary system 
with a separation of 0\farcs304.
They were also detected in the 3.6 $\micron$ and 
4.5 $\micron$ images of the Infrared Array Camera (IRAC)
on the Spitzer Space Telescope, 
although both were not spatially resolved (figure \ref{Spitzer}).
Their magnitudes are $15.8\pm0.6$ mag and $15.6\pm0.9$ mag at
3.6 $\micron$ and 4.5 $\micron$, respectively.
The photometric results for both have unusually large uncertainties.
The bright halo component of the central star as well as diffuse
cirrus emission in the Pleiades cluster dominate the background
around the candidates in both wavelengths.
The companions were not detected in the 5.8 $\micron$ and 8.0 $\micron$ images.
Their combined $K$-band magnitude is $17.5\pm0.2$ mag and their
combined $H$-band magnitude is 17.0 mag.
An X-ray source 1RXH J034628.8+245555 is coincident with the position of
cc1 and cc2.

\begin{figure}[h!!!]
\centering
\includegraphics[width=14.0cm, angle=0]{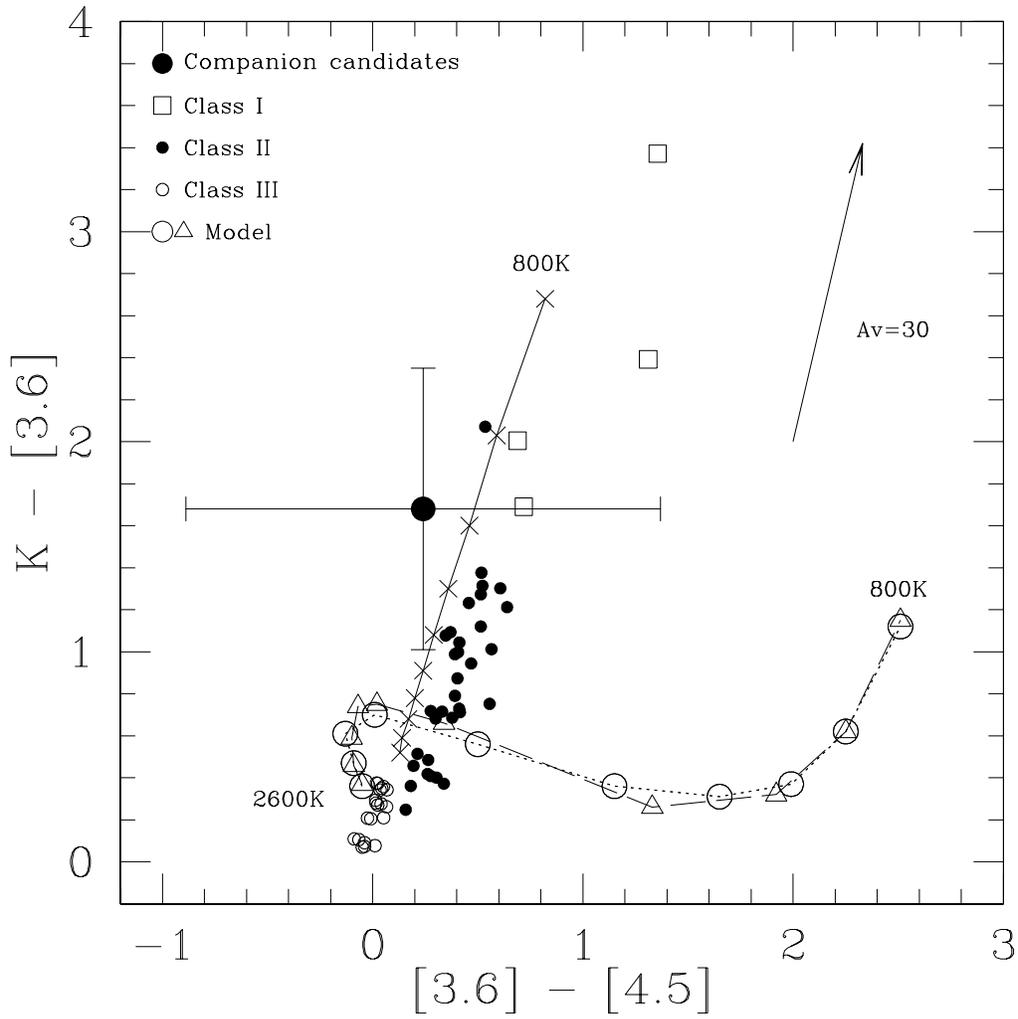}
\begin{minipage}[]{140mm}
\caption{ 
Near-infrared color--color diagram of V1171 Tau/cc1+cc2 (large filled circle).
The open circles and triangles represent the colors of low-effective temperature
objects calculated from the synthesized spectra of
the Cc50 model and the T18c50 model, respectively (Tsuji et al. 2004).
The effective temperatures range from 800 K to 2600 K.
The crosses indicate the colors of blackbodies in the same temperature range.
The small open boxes, the small filled circles, 
and the open circles represent the colors of
Class I, II, and III objects, respectively (Hartmann et al. 2005).
An arrow marks the visual extinction.
\label{cc}
}
\end{minipage}
\end{figure}

Figure \ref{cc} shows a $K$--[3.6]--[4.5] color--color diagram of cc1 and cc2.
The reddening vector was derived using the extinction law of Indebetouw et al.
(2005): $\frac{E_{K-3.6}}{E_{3.6-4.5}}=4.3$ and $E_{3.6-4.5}=0.011A_{V}$.
Colors of heavily reddened ($A_{\rm V}\sim40$ mag)
Class III objects and main-sequence stars 
matched those of the companion candidates.
However, it is unlikely that a dense molecular cloud is associated with
V1171 Tau, as $A_{\rm V}$ is as small as 2 mag toward the Pleiades
cluster (Cernis 1987).
Thus, we rejected the possibility that the companion candidates are heavily 
reddened Class III objects or main-sequence stars.

To investigate the nature of these objects, we calculated
the infrared photospheric colors
of low-mass objects using synthesized spectra (Tsuji et al. 2004).
We used models both with and without dust condensation
for objects whose log $g$ was 5.0.
As seen in figure \ref{cc}, we found that 
no photospheric model spectra matched the observed infrared color.
We also calculated infrared colors of blackbodies.
The companion candidates have colors similar to a 1200 K blackbody.
The expected $H-K$ color of a 1200 K blackbody is 1.7 mag;
however, the observed $H-K$ color of the object was $-0.5$ mag.

The companion candidates are located close to the Class I objects
in the color--color diagram.
We expect that they are Class I objects surrounded by a
circumstellar envelope, and that
higher-temperature photospheric components dominate at wavelengths
shorter than the $K$-band.
We must also consider the possibility that cc1 and cc2 are
background objects, although this is unlikely.
Hempel et al. (2008) found extremely red galaxies whose colors are similar
to those of the companion candidates in the $K$, [3.6], and [4.5] bands.
Such galaxies often form (apparent) binaries.

\begin{figure}[h!!!]
\centering
\includegraphics[width=14.0cm, angle=0]{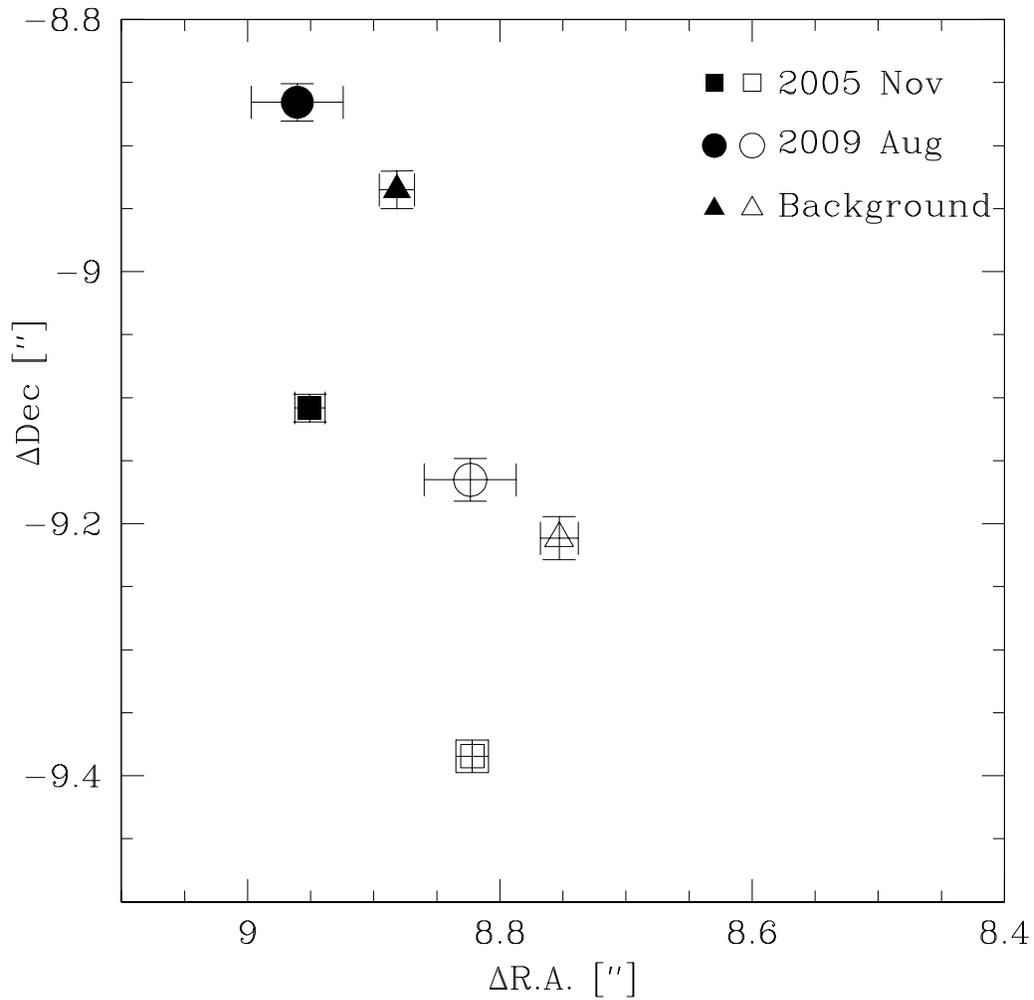}
\begin{minipage}[]{140mm}
\caption{ 
Positions of V1171 Tau/cc1 and cc2 (squares and circles).
Offset positions are shown with respect to V1171 Tau.
The filled and open symbols indicate the positions of cc1 and cc2,
respectively.
The expected positions of background galaxies on 2009 Aug 15 are also
shown (triangles).
\label{pos}
}
\end{minipage}
\end{figure}

A proper motion test can distinguish these two possibilities.
We made this test by using the CIAO data taken in 2005 and the IRCS data taken 
in 2009 (figure \ref{pos}).
If these objects were young companions associated with V1171 Tau,
their positions relative to V1171 Tau would not change during two observational
epochs ($\Delta\alpha_{1}=+8\farcs951, \Delta\delta_{1}=-9\farcs108$ 
for cc1 and $\Delta\alpha_{2}=+8\farcs822, \Delta\delta_{2}=-9\farcs385$
for cc2).
In contrast, if they were background objects, their relative positions 
would change.
Since V1171 Tau has a proper motion of ($\mu_{\alpha}$, $\mu_{\delta}$)
=(+18.4 mas/yr, $-46.2$ mas/yr) (Naval
Observatory Merged Astrometric Dataset catalog), the relative positions
should change by $-0\farcs069$ in R.A. and +0\farcs173 in Dec.
Thus, the expected relative positions were $\Delta\alpha_{1}=+8\farcs882, 
\Delta\delta_{1}=-8\farcs935$ for cc1 and $\Delta\alpha_{2}=+8\farcs753, 
\Delta\delta_{2}=-9\farcs212$ for cc2 regarding the IRCS observations.
The relative positions of the candidates measured in the IRCS data
were $\Delta\alpha_{1}=+8\farcs961, \Delta\delta_{1}=-8\farcs866$ and 
$\Delta\alpha_{2}=+8\farcs823, \Delta\delta_{2}=-9\farcs165$.
These positions were close to the positions predicted under the background
object hypothesis.
We concluded that these faint sources were background objects.

Finally, we did not detect the other faint sources in the Spitzer/IRAC images.
Cc4 is listed in the Automated Plate Machine (APM) catalog, 
where its $b$- and $r$-band
magnitudes are given as 21.13 mag and 19.15 mag, respectively.
Tanner et al. (2007) reported a visual companion of V 1171 Tau with a
separation of 12\farcs7 and a position angle (PA) of 45\fdg0, 
whose $K$-band magnitude was 16.0 mag.
We did not detect such a point source, 
although it should be within the CIAO field of view.

\subsection{HD 282952}

\begin{figure}[h!!!]
\centering
\includegraphics[width=14.0cm, angle=0]{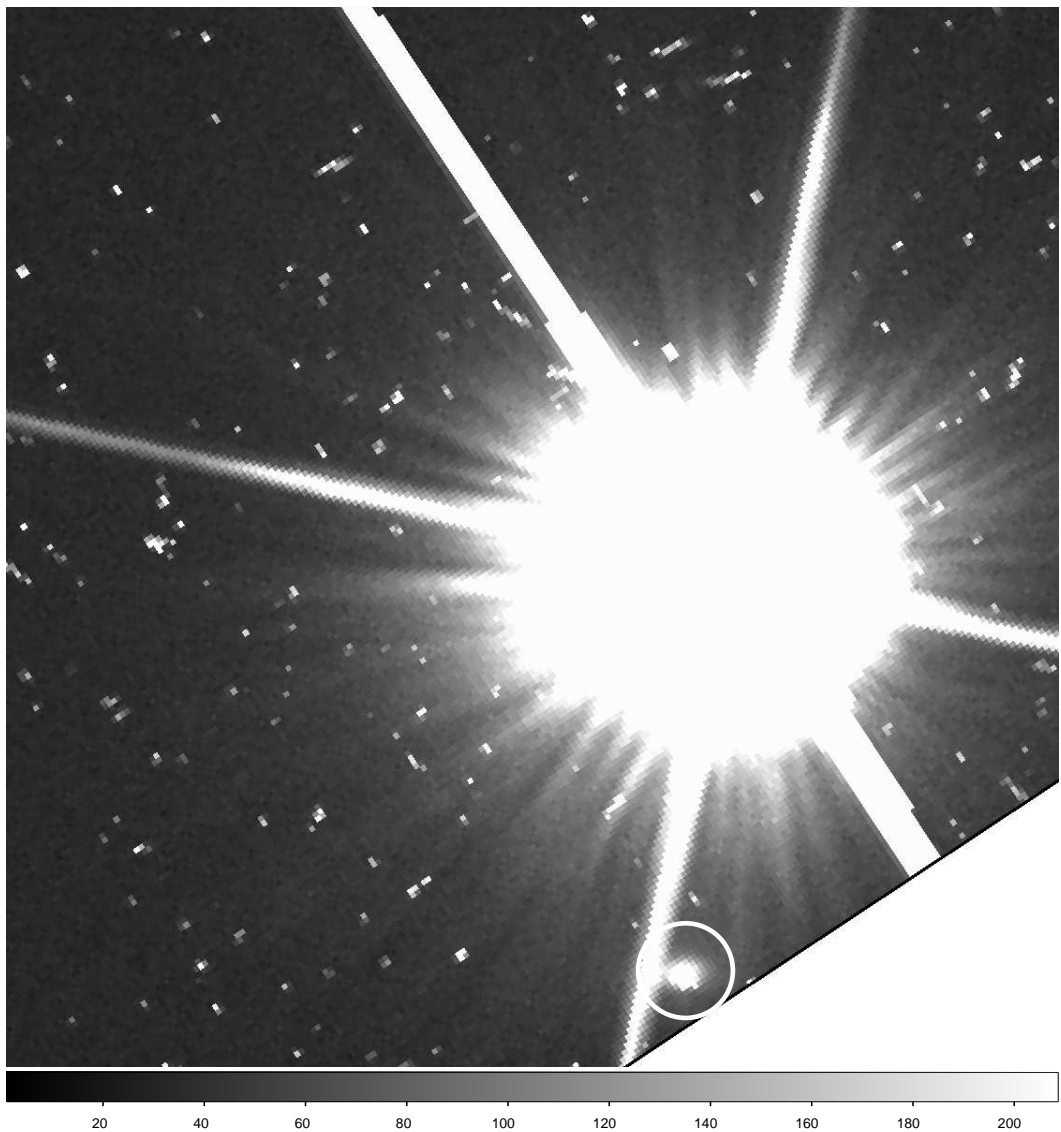}
\begin{minipage}[]{140mm}
\caption{ 
F606W HST image of HD 282952.
The companion candidate is identified by a circle.
The other structures, aside from the central star, are cosmic rays, diffraction patterns,
and saturated pixels.
The field of view is 20\arcsec$\times$20\arcsec.
The lower-right corner is not imaged.
\label{HST}
}
\end{minipage}
\end{figure}

The top-right panel of figure \ref{allfig} shows the $K$-band image of HD 282952. 
One faint source (HD 282952/cc1) was detected, at a separation and PA
from the central star of 7\farcs30 and 174\fdg44, respectively.
This faint source seemed to be extended; its FWHM was 0.3\arcsec.
HD 282952 was also observed by the Hubble Space Telescope (HST) in 1999
with the F606 filter (figure \ref{HST}).
The $V$-band magnitude of the faint source was
23.2 mag.
Its separation and PA at that time
were 7\farcs72 and 173\fdg55, respectively.
If the faint source is physically associated with HD 282952, 
both objects must share a common proper motion
because the orbital motion of the assumed companion 
would be negligibly small in a system with a large semimajor axis.
Thus, a real companion should show no change in separation and 
PA between the epochs of the HST
and Subaru observations.
If the faint source is a background object, 
the separation and PA should vary
as a result of the proper motion of the central star. 
The proper motion of the central star is 
($\mu_{\alpha}, \mu_{\delta}$) = (+18.5 mas/yr, $-48.5$ mas/yr) (Naval
Observatory Merged Astrometric Dataset catalog). 
With these values, we estimated that the separation and PA
should be 7\farcs38 and 174\fdg22 at the time of the CIAO observation, 
if the faint source was indeed a background object.
As described above, the separation and PA at the CIAO observation were
7\farcs30 and 174\fdg44, in relative agreement with the
background object assumption.
We concluded that HD 282952/cc1 is not a faint companion, but a distant
background object.
However, we note that the heavy saturation of HD 282952 itself in the HST image
may degrade the measurement accuracy of the faint-source separation and PA.

\subsection{HD 23269}

The $K$-band image of HD 23269 is shown in the top-left panel of
figure \ref{allfig}.
Due to inclement weather conditions,
wavefront compensation by the AO was poor; the
FWHM of the PSF was 0\farcs3.
We did not use an occulting mask for this target.
A relatively bright companion candidate was detected
at 2\farcs78 west of the central star.
Photometry relative to the central star indicated the magnitude of the 
companion candidate was 12.19 mag.
If this object is associated with the Pleiades cluster, its absolute magnitude
is 6.87 mag at the $K$-band, corresponding to 0.3 M$_{\odot}$ 
and an effective temperature of 3400 K at an age
of 115 Myr (Baraffe et al. 1998).

Kaehler (1999) searched for binaries in the Pleiades cluster using
photometric excess. 
They reported a 0.03 mag excess at the $V$-band for
HD 23269, though this excess is well within the photometric uncertainty.
We believe that the contribution of the candidate to the total $V$-band
magnitude is very small,
since the magnitude difference between the central star and the candidate is
3.68 mag, a luminosity ratio of about 30.
Thus, this discovery of a companion candidate does not contradict
the photometric results of Kaehler (1999).
We could not find the candidate in 2MASS images,
due to their rather poor resolution.

\section{Conclusions}

A near-infrared coronagraphic survey of the Pleiades cluster revealed
10 faint point sources around 7 dwarfs, whose magnitudes were
as faint as 20 mag in the $K$-band.
A pair of companion candidates around V 1171 Tau exhibited an infrared excess,
but the proper-motion measurements indicated that both objects were
distant galaxies.
HD 282952 and its faint companion candidate 
also do not share common proper motions,
suggesting that they are not physically associated with each other.

\bigskip

We are grateful to S. Harasawa for the observations.
This work is partly supported by "The 21st Century COE program: The Origin and
Evolution of Planetary Systems" of the Ministry of Education, Culture,
Sports, Science and Technology (MEXT), Japan.
Y.I. is supported by a Grant-in-Aid for Scientific Research No. 16740256.

\clearpage

\begin{table}[h!!!]
\small
\centering
\begin{minipage}[]{35mm}
\caption[]{Observation log}\label{log}\end{minipage}
\tabcolsep 6mm
 \begin{tabular}{llccc}
  \hline\noalign{\smallskip}
HII & Other name & $K$ mag & Mask diameter & Integration (sec, coadd, frame)\\
  \hline\noalign{\smallskip}
405  & HD 23269  & 8.51 & none     & $10 \times 1 \times 4$ \\
996  & HD 282963 & 8.92 & 0\farcs6 & $10 \times 1 \times 25$ \\
1015 & HD 282952 & 8.99 & 0\farcs6 & $10 \times 1 \times 38$ \\
1032 & V1171 Tau & 9.16 & 0\farcs6 & $10 \times 1 \times 25$ \\
1101 & HD 282954 & 8.76 & 0\farcs6 & $10 \times 1 \times 25$ \\
1139 & HD 23513  & 8.24 & none     & $10 \times 1 \times 34$ \\
1514 & HD 282967 & 8.95 & 0\farcs6 & $10 \times 1 \times 25$ \\
2462 &           & 9.60 & 0\farcs6 & $10 \times 1 \times 25$ \\
2506 & BD +22574 & 8.80 & 0\farcs6 & $10 \times 1 \times 25$ \\
  \hline\noalign{\smallskip}
HII & Other name & $H$ mag & Mask diameter & Integration (sec, coadd, frame)\\
  \hline\noalign{\smallskip}
253  & BD +24541 & 9.12 & 0\farcs6 & $10 \times 3 \times 24$ \\
296  & V966 Tau  & 9.51 & 0\farcs6 & $10 \times 3 \times 24$ \\
1032 & V1171 Tau & 9.27 & 0\farcs6 & $10 \times 3 \times 24$ \\
2462 &           & 9.70 & 0\farcs6 & $10 \times 3 \times 18$ \\
  \noalign{\smallskip}\hline
\end{tabular}
\end{table}

\begin{table}[h!!!]
\small
\centering
\begin{minipage}[]{48mm}
\caption[]{Companion candidates}\label{comp}\end{minipage}
 \begin{tabular}{lcccccl}
  \hline\noalign{\smallskip}
Name & Separation (\arcsec) & Position angle ($^{\circ}$) & $H$ mag & $K$ mag & Aperture (pixels) & Comment \\
  \hline\noalign{\smallskip}
HD 23269/cc1  &  2.78 &  -69.25 &       & 12.2 & 20 \\
HD 282963/cc1 &  9.52 &  -58.39 &       & 18.6 & 8 \\
HD 282952/cc1 &  7.30 &  174.44 &       & 16.9 & 30 & extended? \\
\begin{tabular}{@{}l@{}}
V1171 Tau/cc1 \\
V1171 Tau/cc2 
\end{tabular} &
\begin{tabular}{@{}l@{}}
12.77 \\ 
12.88   
\end{tabular} &
\begin{tabular}{@{}l@{}}
135.50 \\
136.77 
\end{tabular} &
$\Bigr\}$ 17.0 &
\begin{tabular}{@{}l@{}}
17.9 \\
18.8 
\end{tabular} &
\begin{tabular}{@{}l@{}}
15 \\
15
\end{tabular} \\
V1171 Tau/cc3 & 12.87 & -127.73 &       & 19.4 & 15 \\
V1171 Tau/cc4 &  9.25 &  127.05 &       & 20.0 & 15 \\
HD 282954/cc1 &  9.13 &  106.00 &       & 16.6 & 15 & extended? \\
HII 2462/cc1  &  9.97 &   84.93 & 16.6  & 17.2 & 15 & extended? \\
BD+22574/cc1  &  8.44 &   51.82 &       & 18.8 & 15 \\
  \noalign{\smallskip}\hline
\end{tabular}
\end{table}

\newpage

\end{document}